\documentclass{JHEP3}
\usepackage{graphics}
\usepackage{epsfig}

\newcommand{\bea}{\begin{eqnarray}}
\newcommand{\eea}{\end{eqnarray}}
\newcommand{\bi} {\begin{itemize}}
\newcommand{\ei} {\end{itemize}}

\def\half{{\frac{1}{2}}}
\def\nl{\nonumber \\}
\title{Hydrodynamics and the Detection of the QCD Axial Anomaly in Heavy Ion Collisions}

\author{
Boaz Keren-Zur and Yaron Oz\\
Raymond and Beverly Sackler Faculty of Exact Sciences \\
School of Physics and Astronomy \\
Tel-Aviv University, Ramat-Aviv 69978, Israel\\
E-mails: \email{kerenzu@post.tau.ac.il},
  \email{yaronoz@post.tau.ac.il}
}
\abstract{
We consider the experimental implications of the axial current triangle diagram anomaly in a hydrodynamic description of high density QCD. We propose a signal of an enhanced production of spin-excited hadrons in the direction of the rotation axis in off-central heavy ion collisions.}


\preprint{arXiv:1002.0804 [hep-ph]}
\begin{document}
\section{Introduction}
It has been argued that the QCD dense matter in relativistic heavy ion collisions (HIC) exhibits properties
of a collective fluid-like motion with low viscosity to entropy ratio (see e.g. \cite{Luzum:2008cw}). Thus,
relativistic hydrodynamics has become an important analysis tool for HIC.
Relativistic hydrodynamics is formulated in terms of conservation laws of the stress-energy tensor 
and various conserved currents. 

It has been recently revealed that the hydrodynamics description exhibits an interesting effect when a global symmetry current of the microscopic theory is anomalous.
This has been first discovered in the context of the gauge/gravity correspondence \cite{Erdmenger:2008rm,Banerjee:2008th,Torabian:2009qk}.
The Chern-Simons term in the gravity action, which corresponds to having an anomalous
global symmetry current in the dual gauge theory, has been shown to modify the
hydrodynamic current by a term proportional
to the vorticity of the fluid.

At first sight the additional vorticity term seemed to contradict the second law of thermodynamics \cite{Landau}.
This, however, has been resolved by a redefinition
of the entropy current in \cite{Son:2009tf}.
In this work we suggest an experimental signal, which is a consequence
of the anomaly effect. We will consider the effect of the vorticity term as well as that of the gauge fields.  

The major effort in the experimental study of QCD topological effects in the context
of HIC has been focused on charge separation. The origin of this effect is the assumption
that in the deconfined phase of QCD a non-trivial, space-dependent, value for the QCD $\theta$ angle can be generated. In this $P$ violating vacuum a strong magnetic field would induce an electromagnetic current along the magnetic field lines. The experimental signature of this effect is an asymmetry in the charge distribution of the scattered particles in non-central collisions \cite{Kharzeev:2007jp,Fukushima:2008xe}.

In this paper we propose an observable which is charge independent.
The basic idea is that the axial charge density,
in a locally uniform flow of massless fermions,
is a measure of the alignment between the fermion spins.
When the QCD fluid freezes out and the quarks bind to form hadrons,
aligned spins result in spin-excited hadrons.
The ratio between spin-excited and low spin hadron production and its angular distribution
may therefore be used as a measurement of the axial charge distribution.
Due to the short lifetime of high-spin hadrons such as the $\rho$ mesons and $\Delta$ baryons,
we propose to focus on narrow resonances such as $\Omega^-$.
We will predict the qualitative angular distribution and
centrality dependence of the axial charge. Our main proposal is that for off-central collisions we expect enhancement of $\Omega^-$ production along the rotation axis of the collision
(see fig. \ref{fig_prediction}).
\FIGURE{\includegraphics[width=7cm]{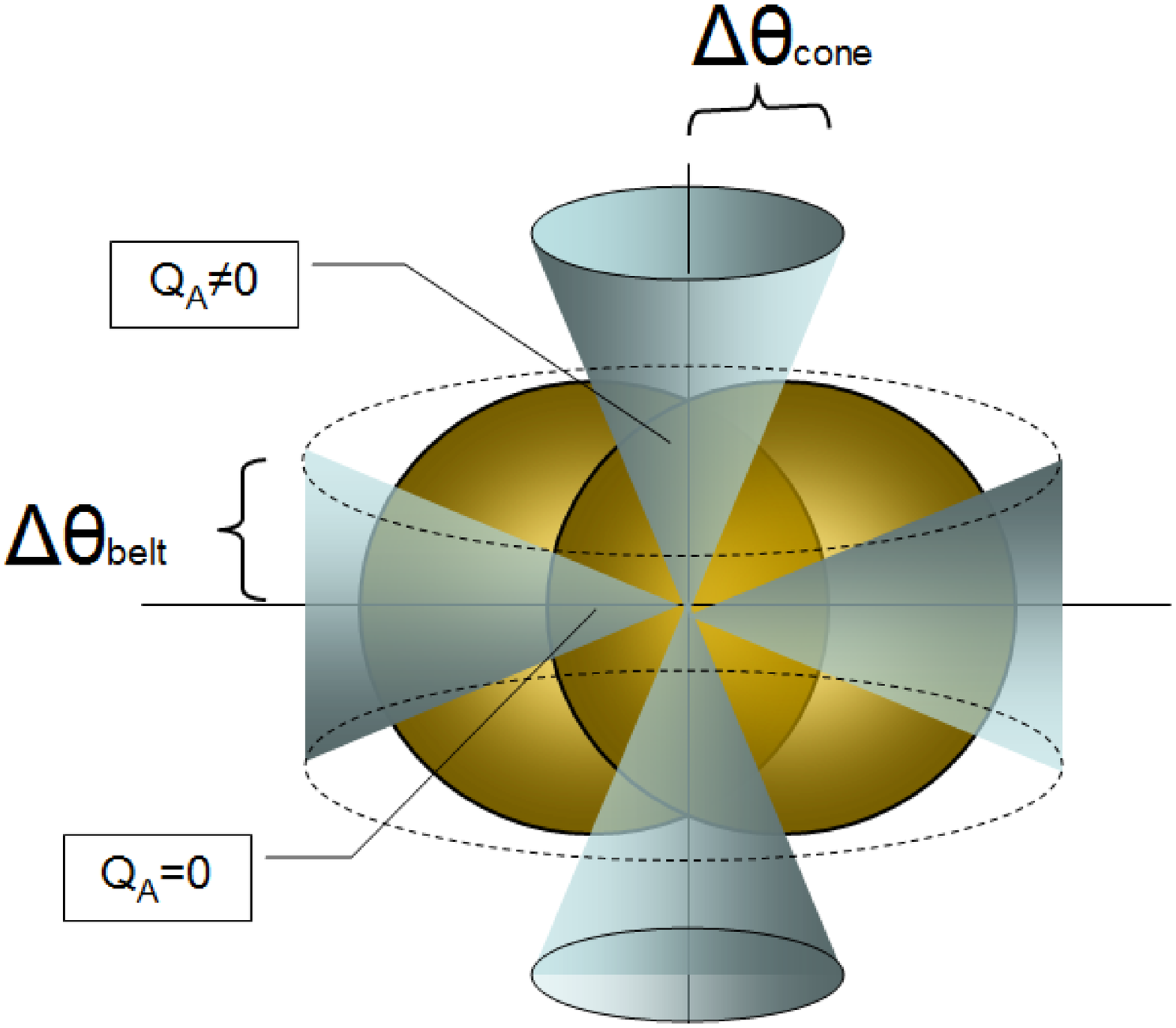}\includegraphics[width=6cm]{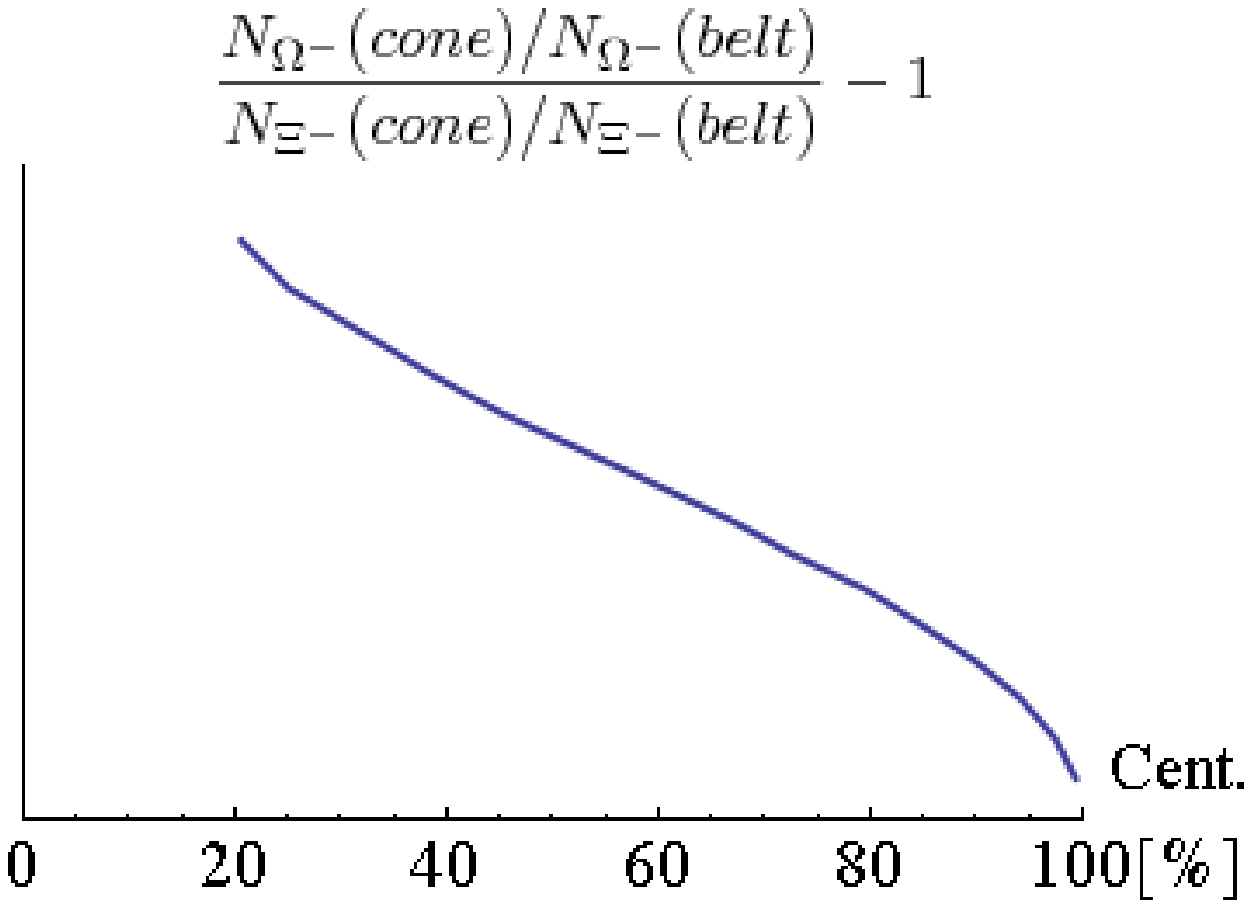}
\caption{The left figure shows an off-central collision of two Gold ions (the beam direction is transverse to the plain of the plot). In blue we see the two spatial angles in which we compare the production rates -- in the upper and lower "cones" we expect to find an enhancement of spin excited hadrons due to the non-zero axial charge $Q_A$
in the fluid, and the "belt" can be used to measure the production without axial charge. The plot on the right shows qualitatively the predicted centrality dependence of the effect.\label{fig_prediction}}}

As we will explain, the calculation of the precise magnitude of the effect requires a detailed numerical analysis as well as making certain assumptions about the hadronization process.
In particular, the Bjorken flow ansatz which is very useful in the numerical analysis of the
hydrodynamics equations cannot be used in this case.
Instead, we will use an estimate for the axial charge distribution at early stages in the evolution of the system.

The paper is organized as follows:
In section \ref{sec_hydro_anomaly} we briefly review the theoretical background for the anomaly effect
in the hydrodynamics framework and discuss the issues involved in detecting this effect in heavy ion collisions.
In section \ref{sec_experimental_signal} we present an
experimental signal, use the Glauber model as the initial condition
to estimate the axial density and give an outline for the data analysis.
The last section is devoted to a discussion and outlook.

\section{Hydrodynamics, Triangle Anomalies and HIC}
\label{sec_hydro_anomaly}

\subsection{Relativistic Hydrodynamics With Anomalous Currents}

The hydrodynamic description of a classical relativistic fluid is a set of conservation equations of
the stress-energy tensor and the global symmetry currents
\begin{equation}
\label{eq_conservation}
\partial_{\mu}T^{\mu\nu} = 0,~~~~~~~\partial_{\mu}j_a^{\mu} = 0 \ ,
\end{equation}
where
\bea
\label{eq_rel_hydro}
T^{\mu\nu}&=&(\epsilon+P)u^\mu u^\nu - g^{\mu\nu}P+\tau^{\mu\nu}\nl
j_a^\mu&=&\rho_au^\mu+\nu_a^\mu \ .
\eea
$u^\mu$ is the fluid velocity field, normalized such that $u_\mu u^\mu=-1$,
$\epsilon$, $P$ and $\rho_a$ are the energy density, pressure and charge densities, respectively.
$\tau^{\mu\nu}$, $\nu_a^\mu$ are
the dissipative terms that contain derivatives of the various fields.
There exists an ambiguity in the definition of the
fields, which we will fix by choosing the Landau frame in which the velocity represents the rest frame of the \emph{energy} density\footnote{
Another useful choice is the Eckart frame, where the velocity is determined by the rest frame of the charge. In this case $\nu_a^\mu=0$.}.
The conditions for the Landau frame are
\bea
u_\mu\tau^{\mu\nu}=u_\mu\nu_a^\mu=0 \ .
\eea

In the Landau frame the global symmetry current takes the form
\bea
j_a^\mu=\rho_a u^\mu-\sigma_a T(g^{\mu\nu}+u^\mu u^\nu)\partial _\nu\left(\frac{\mu_a}{T}\right) + \sigma_a E_a^\mu\ ,
\eea
where $T$, $\mu_a$ and $\sigma_a$ are the temperature, chemical potentials and the conductivities of the medium, and $E_a^\mu\equiv F^{\mu\nu}_au_\nu$ is an external field which is coupled to the current $j_a^\mu$.
This form of the current is modified when
the global symmetry current corresponds to an anomalous current in the microscopic theory \cite{Erdmenger:2008rm,Banerjee:2008th,Torabian:2009qk,Son:2009tf}.
In the latter case the current takes the form
\bea
\label{eq_modified_current}
j_a^\mu&=&\rho_a u^\mu-\sigma_a T(g^{\mu\nu}+u^\mu u^\nu)\partial _\nu\left(\frac{\mu_a}{T}\right)+ \sigma_a E_a^\mu+\xi_a\omega^\mu+\xi^B_{ab}B_b^\mu \ ,
\eea
where the vorticity $\omega^\mu$ and the magnetic field $B_b^\mu$ are defined by
\bea
\omega^\mu&\equiv&\half\epsilon^{\mu\nu\lambda\rho}u_\nu\partial_\lambda u_\rho \nl
B_b^\mu&\equiv&\half \epsilon^{\mu\nu\lambda\rho}u_\nu (F_{b})_{\lambda\rho}~.
\eea

The vorticity and magnetic field coefficients for abelian currents read \cite{Son:2009tf}
\bea
\label{eq_xi_xi_B}
\xi_a&=&C_{abc}\mu_b \mu_c-\frac{2}{3}\rho_aC_{bcd}\frac{\mu_b\mu_c\mu_d}{\epsilon+P} \nl
\xi_{ab}^B&=&C_{abc} \mu_c -\frac{1}{2}\rho_aC_{bcd}\frac{\mu_c\mu_d}{\epsilon+P} \ .
\eea
$C_{abc}$ is the coefficient of the triangle anomaly of the currents $j^\mu_a$,$j^\mu_b$ and $j^\mu_c$,
\bea
C_{abc}&=&\frac{\sum_iQ^i_aQ^i_bQ^i_c}{2\pi^2}\nl
\partial_\mu j^\mu_a&=&-\frac{1}{8}C_{abc}\epsilon^{\mu\nu\sigma\rho}F_{\mu\nu}^bF_{\sigma\rho}^c~.
\eea
and $Q_a^i$ is the charge of the $i$'th Dirac fermion with respect to the $a$ symmetry.
Note that we absorbed the coupling constant in the definition of the gauge fields.

\subsection{Heavy Ion Collisions and the Axial Anomaly}
\label{sec_hydro_rhic}
A set-up in which the hydrodynamics description seems useful is the description of high density
QCD matter created in
heavy ion collisions. In very energetic collisions the hot dense QCD matter can go through a phase transition
into a deconfined phase described by a fluid-like collective motion of quarks and gluons.
The comparison of the relativistic hydrodynamics simulations with the data suggests that
the relativistic fluid is characterized by a low shear viscosity to entropy ratio, which is a property
of strongly coupled systems.
In this work we will relate to experimental observables the effect of triangle anomalies on the hydrodynamics description
discussed in the previous section.

We will consider a deconfined QCD fluid phase, with three light flavors and chiral symmetry restoration.
The global $U(1)$ currents correspond to
$U(1)_B$, the Cartan subalgebra of $SU(3)$ of flavor (which we denote by $U(1)_{I3}$ and $U(1)_S$), and their
axial versions. Accordingly, the relevant currents will be denoted by $j_B^\mu,j_{I}^\mu,j_S^\mu$ and  $j_A^{\mu5},j_{I}^{\mu5},j_S^{\mu5}$. The electromagnetic current will be considered as a linear combination of the vector currents, $j_\gamma^\mu = j_{I}^\mu+\half (j_B^\mu+j_S^\mu)$.
In this work we are interested in the axial current for which the relevant triangle anomalies are

\bea
&&C_{ABB}=\frac{1}{2\pi^2},~~C_{AII}=\frac{3}{4\pi^2},~~C_{ASS}=\frac{3}{2\pi^2},
~~C_{ABS}=-\frac{1}{2\pi^2},~~C_{A\gamma\gamma}=\frac{1}{\pi^2}
\eea
The vorticity coefficient will therefore be given by
\bea
\label{eq_xi}
\xi_A&=&\frac{1}{2\pi^2}\big(\mu_B^2  + \frac{3}{2}\mu_I^2 +3\mu_S^2-2\mu_B\mu_S \big)+\mathcal {O}(\rho_A)
\eea
where, as will be discussed in section \ref{sec_estimate}, we neglect the terms in the vorticity coefficients proportional to the axial density because they are subleading. The only external magnetic field coupled
to these currents is the electromagnetic field, therefore we will use the linear combination of these
global currents which couple to this field
\bea
\label{eq_xi_B}
\xi_{A\gamma}^B
=\frac{1}{\pi^2}\mu_\gamma+\mathcal {O}(\rho_A)
=\frac{1}{2\pi^2}\big(2\mu_I+\mu_B+\mu_S\big)+\mathcal {O}(\rho_A)
\eea

In the deconfined phase there are in principle additional degree of freedom that must be taken into
account, and these are the color current and gluon fields.
We consider these as external sources and their contribution to the
equations of motion for the axial current is in the (non-)conservation equation for the axial current
\bea
\partial_\mu j^\mu_A&=&-\frac{1}{8}C_{A\gamma\gamma}\epsilon^{\mu\nu\sigma\rho}F_{\mu\nu}F_{\sigma\rho}
-\frac{1}{8}C_{ACC}\epsilon^{\mu\nu\sigma\rho}G^a_{\mu\nu}G^a_{\sigma\rho}~.
\label{axialcurrent}
\eea
where $G^a_{\mu\nu}$ is the gluon field, and
\bea
C_{ACC}=\frac{3}{4\pi^2} \ .
\eea
We will estimate this effect, but will not assign a nonzero chemical potential to the color current.

Note that equation (\ref{axialcurrent}) requires some clarification. In a hydrodynamics description one considers conserved currents,
while the axial current is not conserved. One can allow a non-conservation of the currents due to external sources in a hydrodynamics framework.
However, can we actually consider the electromagnetic and gluon gauge fields as external sources?
The fact that fields generated by the fluid are considered here as external, is similar to the standard discussion of stellar magneto-hydrodynamics, where the magnetic fields generated by the rotation of the star are considered as an external force. Moreover, in an off-central collision the gauge fields are generated mostly by spectator nucleons, and can be treated as external force terms.

When attempting to make precise quantitative predictions one encounters several issues.
First, we need to estimate the chemical potentials appearing in the formula for the coefficients $\xi_A$ (\ref{eq_xi}) and (\ref{eq_xi_B}). We will provide such an estimate in \ref{sec_estimate}. In general, in the analysis of HIC in the hydrodynamic framework one assumes a small chemical potential compared to the temperature, which can imply an effect too small to be detected.
Second, in most analyses of HIC, one assumes a Bjorken flow ansatz, where all the observables are boost invariant.
This means that the divergences of the vorticity and magnetic fields are zero, and the equations for the axial density are trivial (see eq. (\ref{eq_conservation}) and (\ref{eq_modified_current})). Another way to see this is to note that the Bjorken ansatz assumes zero angular momentum in the reaction plane,
while the existence of large angular momentum is the main source for vorticity and magnetic field in the fluid.
Solving the hydrodynamics equations without the Bjorken ansatz is a difficult task, which we will leave for future work.
Instead we will estimate the axial density distribution at an early stage of the flow without fully solving
the complete set of equations.
Third, there is no direct experimental access to the axial charge distribution of the fluid, since
all the information regarding chirality is erased during the hadronization process. We will propose a solution to this problem in the next section.

\section{An Experimental Signal}
\label{sec_experimental_signal}
\subsection{The Axial Charge and Enhanced Production of High-Spin Hadrons }
\label{sec_signal}

The phase transition from a fluid state of QCD matter into hadron gas is arguably the least understood stage in the hydrodynamic description of the collision. It is unclear where and when the phase transition occurs and how exactly the free quarks bind and form hadrons.
Nevertheless, using a phenomenological description of the process we argue that non-zero axial charge in the context of heavy ion collision can lead to an enhancement in the
production of spin-excited hadrons.

For simplicity, we will assume that the momenta of the fermions in a small volume of moving fluid are pointing in the same direction (see figure \ref{fig_cone_alignedspins}).
Note also,  that in the zero mass limit, a non-zero axial charge means a preferred helicity for these fermions.
The combination of these two statements means that a non-zero axial charge enhances the probability for quark spins to be aligned.
When the fluid freezes-out and particles with aligned spins bind to form hadrons,
the bound states cannot have low intrinsic spin\footnote{for example, a pseudo-scalar meson can only be formed by quarks with anti-aligned spins}. We therefore propose
that a non-zero axial charge enhances the production rate of spin-excited hadrons.

In order to make this more quantitative, let us define
\bea
\lambda_{i}(\Omega)   \equiv\frac{N_{i} (\Omega)   }{N_{tot} (\Omega)}~,
\eea
where $N_i(\Omega)$ is the number of hadrons of species $i$
detected in a solid angle $\Omega$ and $N_{tot} (\Omega)$ is the total number of detected particles in that angle.
In order to find the dependence of this quantity on the axial charge in a volume element, $Q_A=\int dV \rho_A=N_L-N_R$, we define
\bea
\lambda^0_{i}   \equiv\frac{N^0_{i} (\Omega)   }{N^0_{tot} (\Omega)},\qquad
\lambda^{*0}_{i}   \equiv\frac{N^0_{i} (\Omega)   }{N^{*0}_{tot} (\Omega)}
\eea
where $N^{*}_{tot} (\Omega)$ is the total number of spin-excited hadrons detected in $\Omega$,
and the index $0$ means that this quantity is evaluated at zero axial charge in the QCD fluid.
Note that assuming a radial flow, the relevant volume of fluid is a cone covered by the angle $\Omega$ (See fig. \ref{fig_cone_alignedspins}).
$\lambda^0_{i}$ and $\lambda^{*0}_{i}$ are parameters which depend on the hadronization process.
They are unknown theoretically and will be measured in scattering angles which cover cones with no axial charge,
\bea
\lambda^0_{i}   =\lambda_{i}(\Omega_{(Q_A=0)})~.
\eea

\FIGURE{\includegraphics[width=12cm]{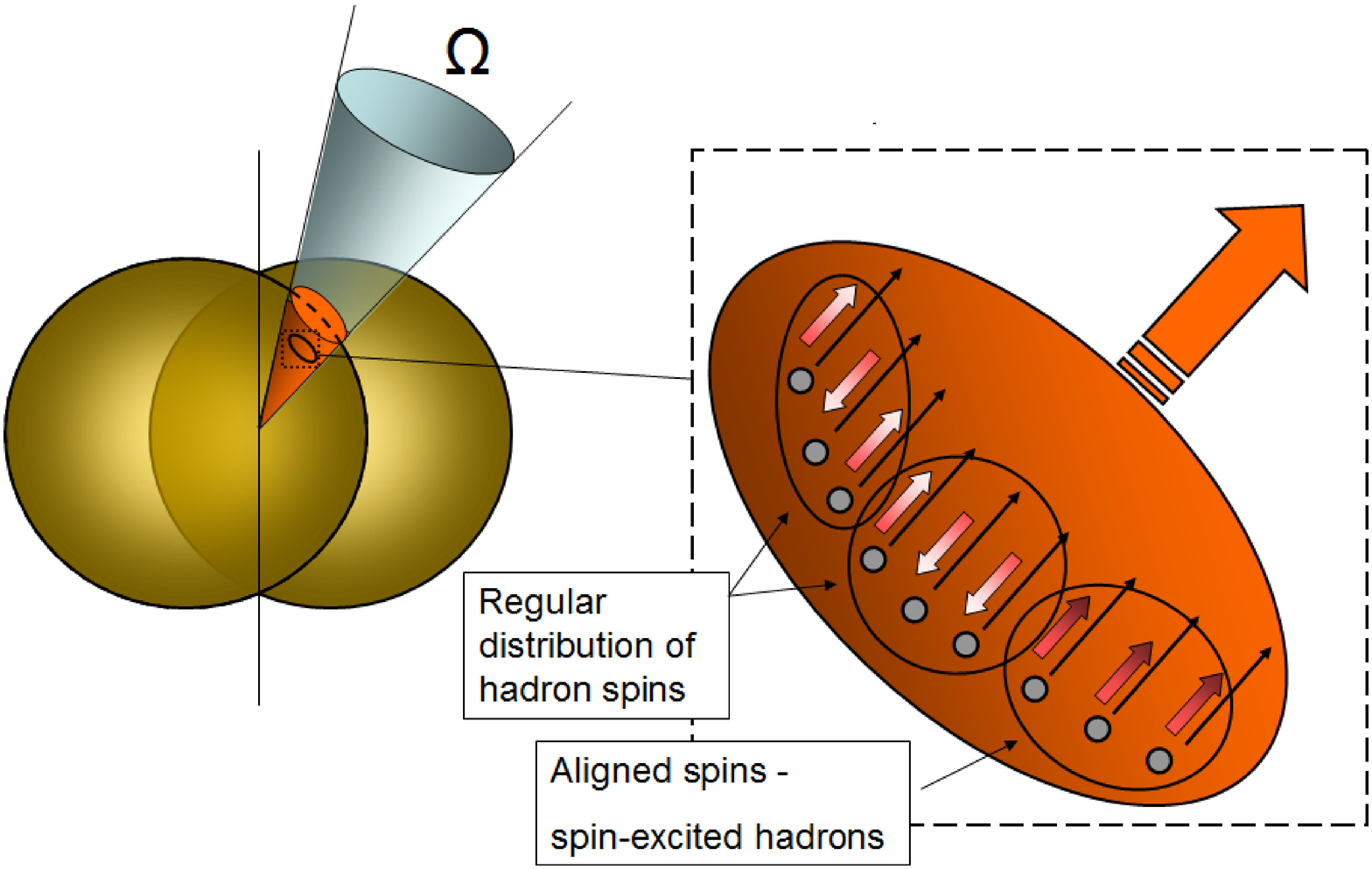}
\caption{The figure on the left shows the spatial angle $\Omega$ and the cone of QCD matter which flows in its direction. On the right we show a zoomed-in cartoon of a small volume of fluid with non-zero axial charge. The small circles represent the fermions and the thick arrows represent their spins. The larger circles represent the bound states. Bound states with aligned spins can form only spin excited hadrons.\label{fig_cone_alignedspins}}}

We now divide the total number of fermions denoted by $n_{tot}$ (both left handed and right handed quarks and anti-quark), in a given volume of QCD fluid into two groups: one with an equal number of left and right fermions, and the other with left handed only (or right handed only, if the axial charge is negative).
Thus, their relative portions in a given volume are $\frac{n_{tot}-|Q_A|}{n_{tot}}$ and $\frac{|Q_A|}{n_{tot}}$ respectively.
While the species and spin of hadrons produced from the first group will be distributed according to the "regular" ratios $\lambda^0_{i}$ as dictated by
the hadronization process,
the second group can only bind into spin-excited hadrons, because the spins are aligned. The species of the particles in the second group, therefore, will be determined by $\lambda^{*0}_{i}$. If we take, for example, the proton as a representative of the low spin hadrons and the $\Delta$ resonance as a representative of the spin-excited hadrons we can write
\bea
\lambda_p (\Omega)      &=&\frac{n_{tot}(\Omega)-|Q_A(\Omega)|}{n_{tot}(\Omega)} \lambda^0_{p}\nl
\lambda_\Delta(\Omega) &=&\frac{n_{tot}(\Omega)-|Q_A(\Omega)|}{n_{tot}(\Omega)} \lambda^0_{\Delta} +\frac{|Q_A(\Omega)|}{n_{tot}(\Omega)}\lambda^{*0}_{\Delta}
\eea
where the notations $n_{tot}(\Omega)$ and $Q_A(\Omega)$ mean the fermion number and axial charge in the
volume covered by the angle $\Omega$.

Using this result we get the following ratio:
\bea
\frac{\lambda_\Delta(\Omega)}{\lambda_p(\Omega)}
&=&
\frac{\lambda^0_{\Delta}}{\lambda^0_{p}}\left(1+\frac{|Q_A(\Omega)|}{n_{tot}(\Omega)-|Q_A(\Omega)|}\frac{\lambda^{*0}_{\Delta}}{\lambda^0_{\Delta}}\right)
\eea
Treating $\frac{\lambda^{*0}_{\Delta}}{\lambda^0_{\Delta}}$ as some unknown parameter of order 1\footnote{$\frac{\lambda^{*0}_{\Delta}}{\lambda^0_{\Delta}}=\frac{N^0_{tot}}{N^{*0}_{tot}}$
 can naively be approximated as $\frac{10}{7}$, using the degeneracy of the various spin multiplets.}, and writing in terms of experimental quantities, we get
\bea
\frac{N_\Delta(\Omega)/N_\Delta(\Omega_{(Q_A=0)})}{N_p(\Omega)/N_p(\Omega_{(Q_A=0)})}-1
&\propto&\frac{|Q_A(\Omega)|}{n_{tot}(\Omega)-|Q_A(\Omega)|}\approx\frac{|Q_A(\Omega)|}{n_{tot}(\Omega)}\propto\frac{|Q_A(\Omega)|}{N_{part}(\Omega)}
\eea
In the last step we assumed that $n_{tot}$, the number of particles in the fluid,
is proportional to the number of nucleons participating in the collision, $N_{part}$, which will be discussed in more detail in the next section.
A more precise analysis of the effect, which requires a numerical solution of the hydrodynamics equations, is suggested in the discussion.

\subsection{The Glauber Model and the Distribution of Axial Charge at Early Stages}
\label{sec_axial_Glauber}
In this section we describe a method of estimating the axial charge distribution at early stages of
the evolution of the system. We focus on the QED contribution, and relate it to QCD effects in the next section.
The non-conservation equation for the axial current in the Landau frame is
\bea
\partial_\mu j_A^\mu =\partial_\mu \big(\rho_A u^\mu-\sigma_A T(g^{\mu\nu}+u^\mu u^\nu)\partial _\nu\left(\frac{\mu_A}{T}\right)+\xi_A\omega^\mu+\xi^B_{A\gamma} B^\mu\big)=C_{A\gamma\gamma}E_\mu B^\mu~,
\eea
where the coefficients $\xi_A$ and $\xi^B_{A\gamma}$ are taken from (\ref{eq_xi}) and (\ref{eq_xi_B}), and
since there is no external field associated with the axial charge, we dropped the $E_A^\mu$ term.
$E_\mu$ and $B^\mu$ are the electromagnetic fields generated in the HIC.

Under the assumption of uniformly distributed chemical potentials and anomaly coefficients
$\partial_\mu \mu_A|_{t_0}\approx\partial_\mu \xi_A|_{t_0}\approx0$, the equation takes the form
\bea
\label{eq_conservation_and_source}
\partial_\mu (\rho_A u^\mu)
= C_{A\gamma\gamma}E_\mu B^\mu-(\xi_A \partial_\mu\omega^\mu+\xi^B_{A\gamma} \partial_\mu B^\mu) \ .
\eea
Thus, the RHS can be considered as source terms for the "classical" axial current, $\rho_Au^\mu$.
At $t=t_0$, the axial density is zero and so is its spatial derivative ($\rho_A=\partial_i\rho_A=0$) and we get
\bea
\partial_t \rho_A|_{t_0}=\frac{1}{u^t}(C_{A\gamma\gamma}E_\mu B^\mu-\xi_A \partial_\mu\omega^\mu-\xi^B_{A\gamma} \partial_\mu B^\mu)|_{t_0}~.
\eea
Given the time derivative of the axial density we can estimate its distribution at early stages
\bea
\rho_A|_{t_0+\Delta t}\approx \Delta t\partial_t \rho_a|_{t_0} = \frac{\Delta t}{u^t}(C_{A\gamma\gamma}E_\mu B^\mu-\xi_A \partial_\mu\omega^\mu-\xi^B_{A\gamma} \partial_\mu B^\mu)|_{t_0}~,
\eea
where $\Delta t$ is a time interval in which this linear approximation is assumed to be valid.

The important assumptions that have been made so far are the uniform distribution of the chemical potentials and
anomaly coefficients, and that for sufficiently short times, the evolution in time of the system can be approximated
by a linear expansion.
We will now estimate the source terms at the initial conditions using the Glauber model.
In this model the energy and velocities at the initial conditions will be obtained assuming that the nucleus density is given by the Woods-Saxon distribution
\bea
\rho(r)=\frac{\rho_0}{1+e^{-(r-R_0)/a_0}}~,
\eea
where $\rho(r)$ is the nucleon density, and for gold ions we use the values $a_0=0.54$ fm and $R_0=6.4$ fm \cite{Luzum:2008cw}. $\rho_0$ is determined by the condition $\int dV \rho = A = 197$.
It is useful to define the "Thickness function"
\bea
T(x,y)=\int_{-\infty}^{\infty}dz\rho(\sqrt{x^2+y^2+z^2})~.
\eea
In a non-central collisions, we set the origin of our coordinate system between the centers of the two colliding ions, and set the $y$ axis along the rotation axis.
The number of participating nucleons, $N_{part}$, is given by
\bea
N_{part}(b)=\int dx dy \Bigg[&~&T(x+\frac{b}{2})\left(1-\left(1-\frac{\sigma_{NN}T(x-\frac{b}{2})}{A}\right)^A\right)\nl
&+&T(x-\frac{b}{2})\left(1-\left(1-\frac{\sigma_{NN}T(x+\frac{b}{2})}{A}\right)^A\right)\Bigg]~,
\eea
where the nucleon-nucleon scattering cross section $\sigma_{NN}\sim40$ mb.
The energy density of the fluid in the Glauber model is proportional to the product of the thickness functions
\bea
\epsilon(x,y)|_{t_0}\propto T(x-b/2)T(x+b/2)~.
\eea

The initial velocity of an infinitesimal area of fluid $v_z(x,y)$ is assumed to be the center of mass velocity at that location
\bea
v_z(x,y)|_{t_0}&\approx&\frac{ \beta T(x+b/2,y)-\beta T(x-b/2,y)}{T(x+b/2,y)+T(x-b/2,y)}
\eea
$\beta\approx1$ is the velocity of the colliding ions. The $z$-component of the 4-velocity, $u_z$, is given by $\gamma v_z$.

We now use this model to evaluate the source terms in (\ref{eq_conservation_and_source}).
We have
\bea
\label{eq_source_vorticity}
\partial_\mu\omega^\mu=\half\epsilon^{\mu\nu\rho\sigma}(\partial_\mu u_\nu)(\partial_\rho u_\sigma)~.
\eea
Since at $t=t_0$ all the velocities are along the $z$ axis, namely in the beam direction,
we find that $u_x|_{t_0}=u_y|_{t_0}=0$, and $u_t|_{t_0}=\sqrt{1+u_z^2}$.
Therefore, at this stage the only non-zero contribution to the divergence of the vorticity is
\bea
\partial_\mu\omega^\mu|_{t_0}=\left[(\partial_t u_x)(\partial_y u_z)-(\partial_t u_y)(\partial_x u_z)\right]|_{t_0}~.
\eea

In order to estimate $\partial_t u_x$ we use the hydrodynamics equations for a perfect fluid:
\bea
u^\nu\partial_\nu u_\mu = \frac{(g_{\mu\nu}+u_\mu u_\nu)\partial^\nu P}{\epsilon + P}
\eea
using the equation of state for conformal hydrodynamics $\epsilon=3 P$ (this condition can be easily relaxed), and the initial conditions $u_x=u_y=\partial_zu_x=0$, we are left with
\bea
\partial_t u_x|_{t_0} = \frac{1}{4\sqrt{1+u_z^2}}\frac{\partial_x \epsilon}{\epsilon}|_{t_0}~.
\eea

Similarly, we can find the divergence of the 4-dimensional magnetic field
\bea
\label{eq_source_magentic}
\partial_\mu B^\mu|_{t_0}&=&\half\epsilon^{\mu\nu\sigma\rho}(\partial_\mu u_\nu) F_{\sigma\rho}|_{t_0}\nl
&=&\Big(F_{xz}(\partial_yu_t-\partial_tu_y)-F_{yz}(\partial_xu_t-\partial_tu_x)-F_{xt}\partial_yu_z+F_{yt}\partial_xu_z\Big)\big|_{t0}~.
\eea
The electromagnetic fields generated by the colliding ions can be obtained by boosting the electromagnetic field from the ion rest frame. The electric field can be estimated assuming the charge density is also given by the Woods-Saxon distribution.

\subsection{Estimating the Magnitude of the Effect}
\label{sec_estimate}

There are several additional factors that we have to estimate:
\begin{itemize}
\item
As discussed in section \ref{sec_signal}, the magnitude of the signal depends on the ratio between the number of fermions in the
fluid and the number of participating nucleons. The number of participating nucleons is $O(100)$, while for having a fluid-like collective motion one needs $O(1000)$ particles, we therefore estimate the required ratio as $O(10)$. 

\item
We take  $\mu_B,\mu_I,\mu_S\sim10$ MeV$\sim0.05 fm^{-1}$ \cite{Andronic:2005yp}

\item
The anomaly coefficient is the sum of two terms
\bea
\xi\propto \mu^2\left(1-\frac{2}{3}\frac{\mu\rho}{\epsilon+P}\right)~.
\eea
The second term can be neglected because in high energy HIC the chemical energy ($\mu\rho$)
is negligible with respect to the total energy.
Also, the axial density is zero at $t_0$ and therefore this term will not effect the initial conditions.

\item
The quantity that we have to compute is the axial density at freeze-out.
This could be obtained using the linear approximation described above only if the
time of freeze-out were in the regime of validity of this linearization.
Since the linear approximation is not valid for the entire process, one must solve the full set of 
equations
for the velocities, energy densities and magnetic fields, and then use them as input for the axial
current EOM. This task will be left for future analysis.
In the following we will assume that the general trend (the axial charge being concentrated
along the rotation axis) will remain the same during the evolution of the system.
This assumption may be supported by the fact that all the terms in eqs. (\ref{eq_source_vorticity}) and
(\ref{eq_source_magentic}) may be reduced due to dissipative effects, but will not change signs.
We will approximate the freeze-out time by $\Delta t \approx  10^{-22}sec = 30$ fm/c.

\item
The boost factor in HIC collisions is taken to be $\gamma=100$.

\item
When integrating over the volume we assume a thickness of order $\frac{R0}{\gamma}$.

\end{itemize}

Before showing the numerical results, let us compare the three QED source terms for the axial current:
\bea
CE_\mu B^\mu &\sim& C\gamma F\widetilde F \sim C\gamma\left(\frac{e^2Z}{R_0^2} \right)^2\nl
\xi^B \partial_\mu B^\mu &\sim& C \mu\gamma F  \partial_t u \sim C \mu\gamma\left(\frac{e^2Z}{R_0^2} \right)\frac{1}{ R_0}\nl
\xi \partial_\mu \omega^\mu &\sim& C  \mu^2\partial_x u \partial_t u \sim C\mu^2\frac{1}{ \zeta R_0^2}
\eea
(where $\zeta$ is a small factor that takes into account that the relevant regions are close to the rotation axis, and not at a distance $R_0$).
We see that the first is larger than the second by a factor of $\frac{e^2Z\zeta}{\mu R_0}\sim 25$, and by a factor of  $\gamma\zeta\left(\frac{e^2Z}{\mu R_0}\right)^2\sim\zeta\cdot 10^5$ than the third.
The dominant contribution will therefore come from the $E\cdot B$ term.
It is worth mentioning that although the dominant terms are linearly proportional to the boost factor $\gamma$,
the volume of integration is inversely proportional to $\gamma$, and therefore the total axial charge should be independent of the collision energy.

The QCD contribution of the chromo-magnetic field is much more difficult to estimate. We will therefore
assume that in the deconfined phase, the dominant
contribution to the colored interaction is the 1-gluon exchange, and that the interaction is similar to the electromagnetic Coulomb interaction, up to a change of coupling constants and group theory factors. In this case, the contribution of the chromo-magnetic
field to the source term of the axial charge is similar in its spatial distribution to the electromagnetic one.
Assuming that this method of finding the chromo-magnetic field is valid,
the ratio between the external chromo-magnetic and magnetic field contributions is $\frac{\alpha_s}{\alpha}\sim 100$.
Finally, note that we are not considering topological effects that can induce a change in the total axial charge.
In the whole discussion, the total axial charge is zero, and we only study the implications of its distribution.

\subsection{Numerical Results}
\label{sec_numerical}
In the following plots we show the numerical results of this analysis for the QED effects.
As discussed above, we do not have precise values for the various anomaly coefficients and
external fields and therefore we cannot accurately add the various contributions. 
However, the general features of the effect (axial charge distribution, centrality dependence)
are similar for all types of contributions, thus combining these results will affect
only the over-all magnitude.

In plot \ref{fig_rhoA} we see the resulting axial charge density as a function of location in the plane transverse to the beam direction.
We see that the areas of largest charge density are located along the axis of angular momentum,
and that along the $x$-axis the charge is zero.
\FIGURE{\includegraphics[width=4cm]{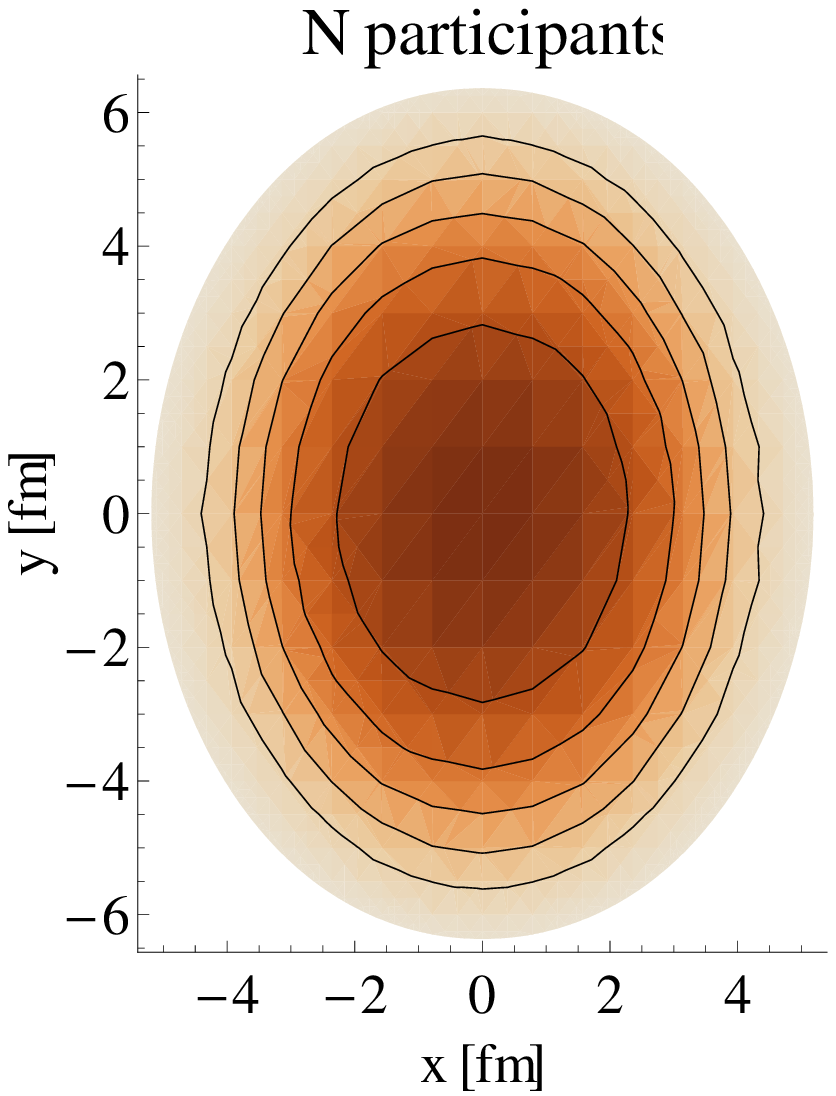}\includegraphics[width=4cm]{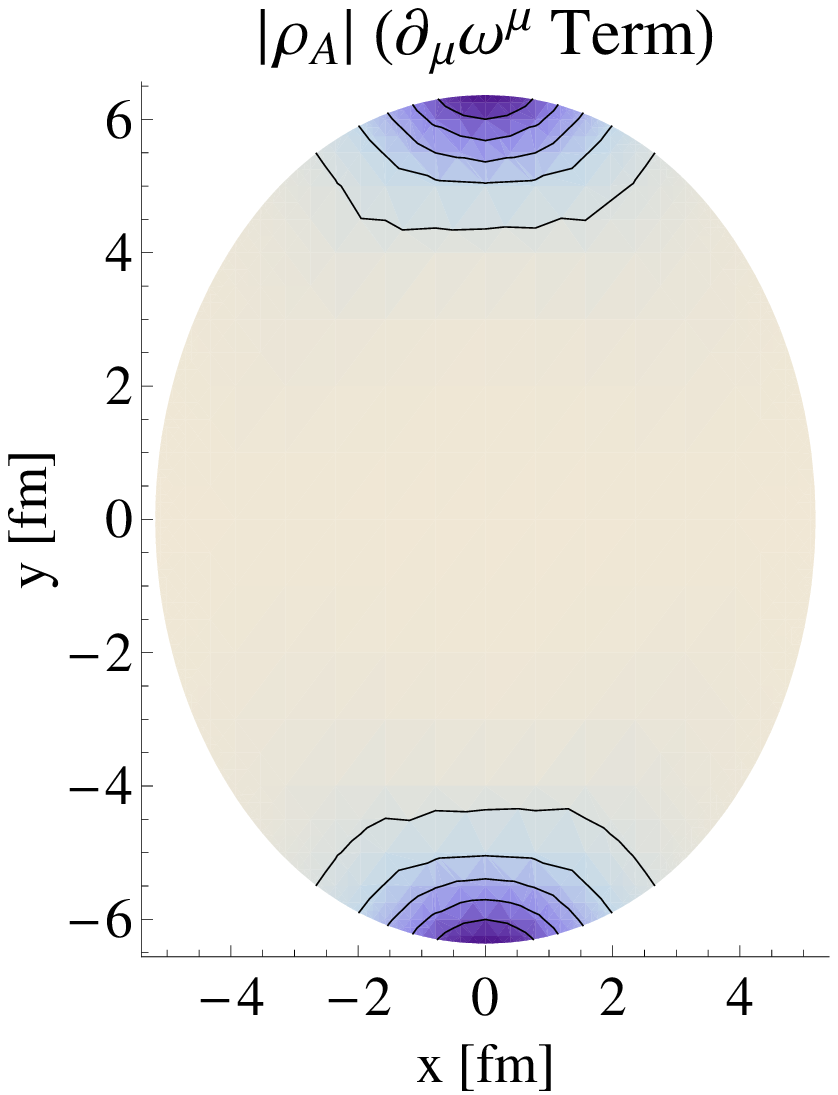}\includegraphics[width=4cm]{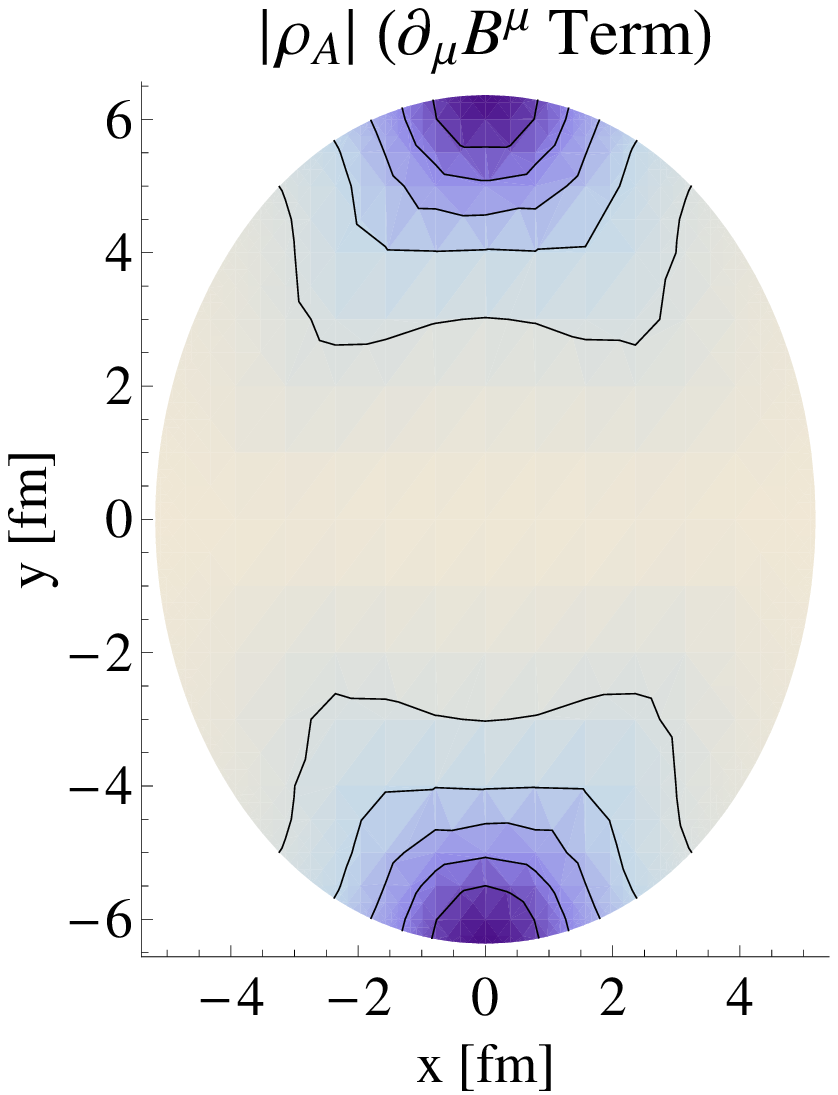}\includegraphics[width=4cm]{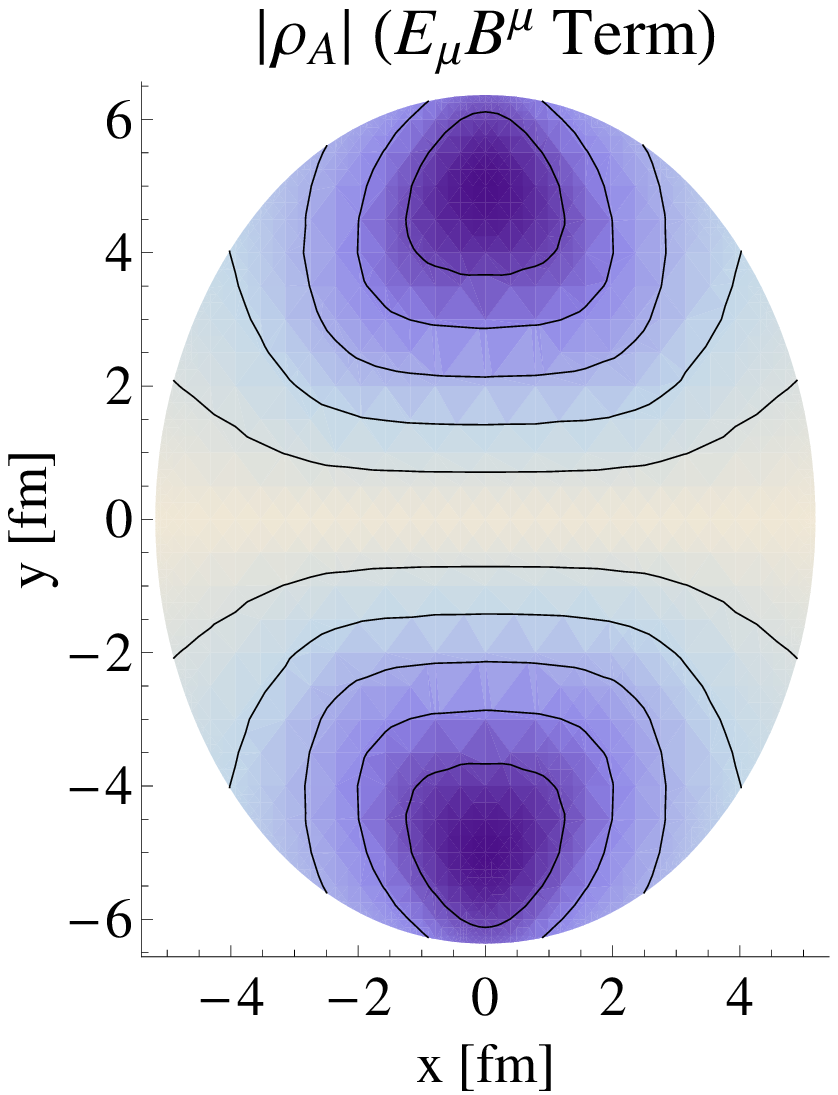}
\caption{The number of participants and axial charge density (dark shade means larger absolute values) at $t=t_0+\Delta t$ for a midcentral collision ($b=R_0$).\label{fig_rhoA}}}
Plots \ref{fig_rhoAx0}, \ref{fig_intrhoA} and \ref{fig_intrhoA_np} demonstrate the dependence of the density on the centrality. As expected, the effect is small for central collisions, because of the low angular momentum.

An important feature of the axial charge distribution is the fact that it is concentrated along the rotation axis.
In plot \ref{fig_pi_2_phi} we show the second moment of the angular distribution
\bea
\label{eq_pi_2_phi}
\langle|\pi/2-\phi|\rangle\equiv\sqrt{\frac{\int dxdy\rho_A(x,y)\arctan^2\frac{x}{y}}{\int dxdy\rho_A(x,y)}}~,
\eea
and its dependence on the centrality ($\phi$ is defined as the angle with respect to the $x$-axis). As will be discussed below, this parameter is relevant for the
detectability of the proposed signal.

\FIGURE{\includegraphics[width=5cm]{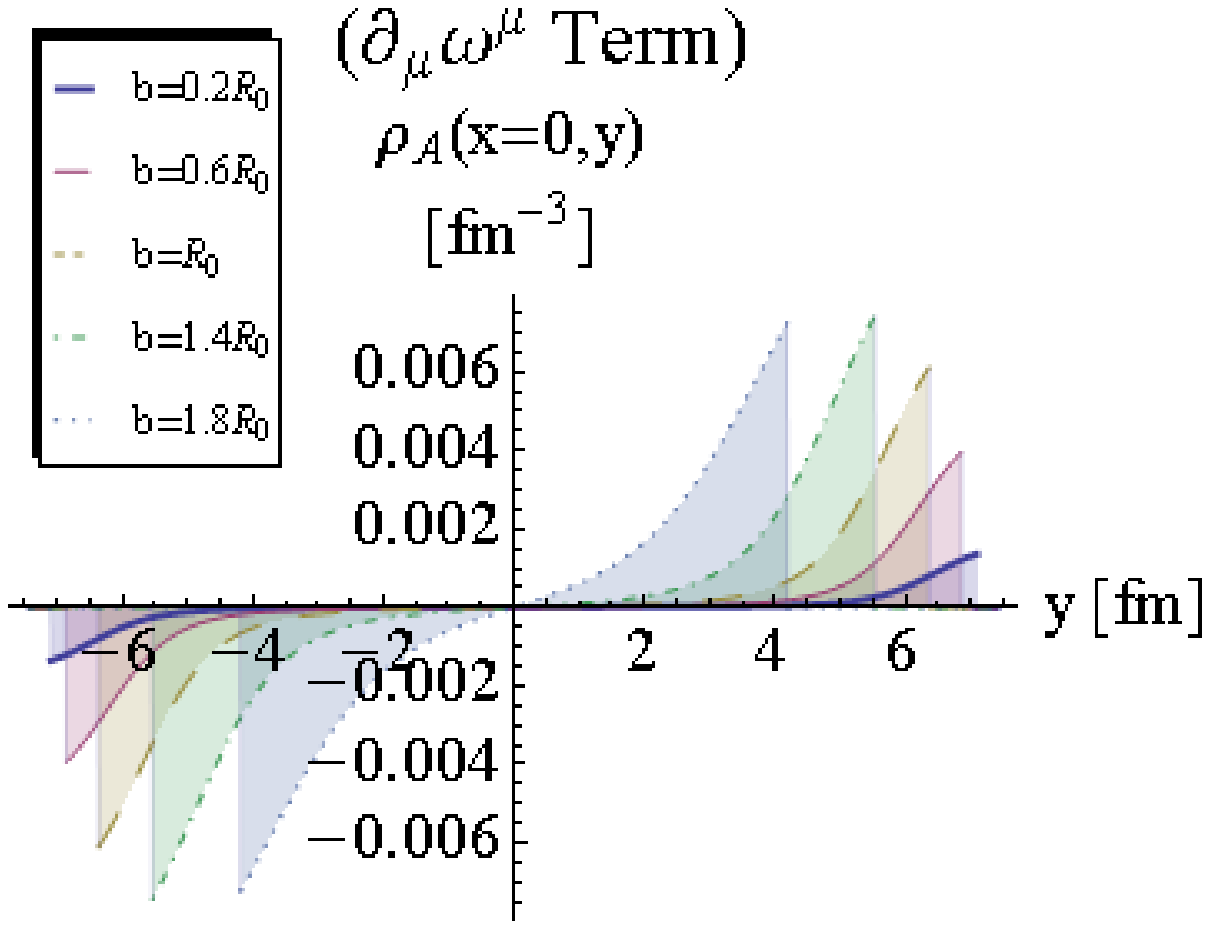}\includegraphics[width=5cm]{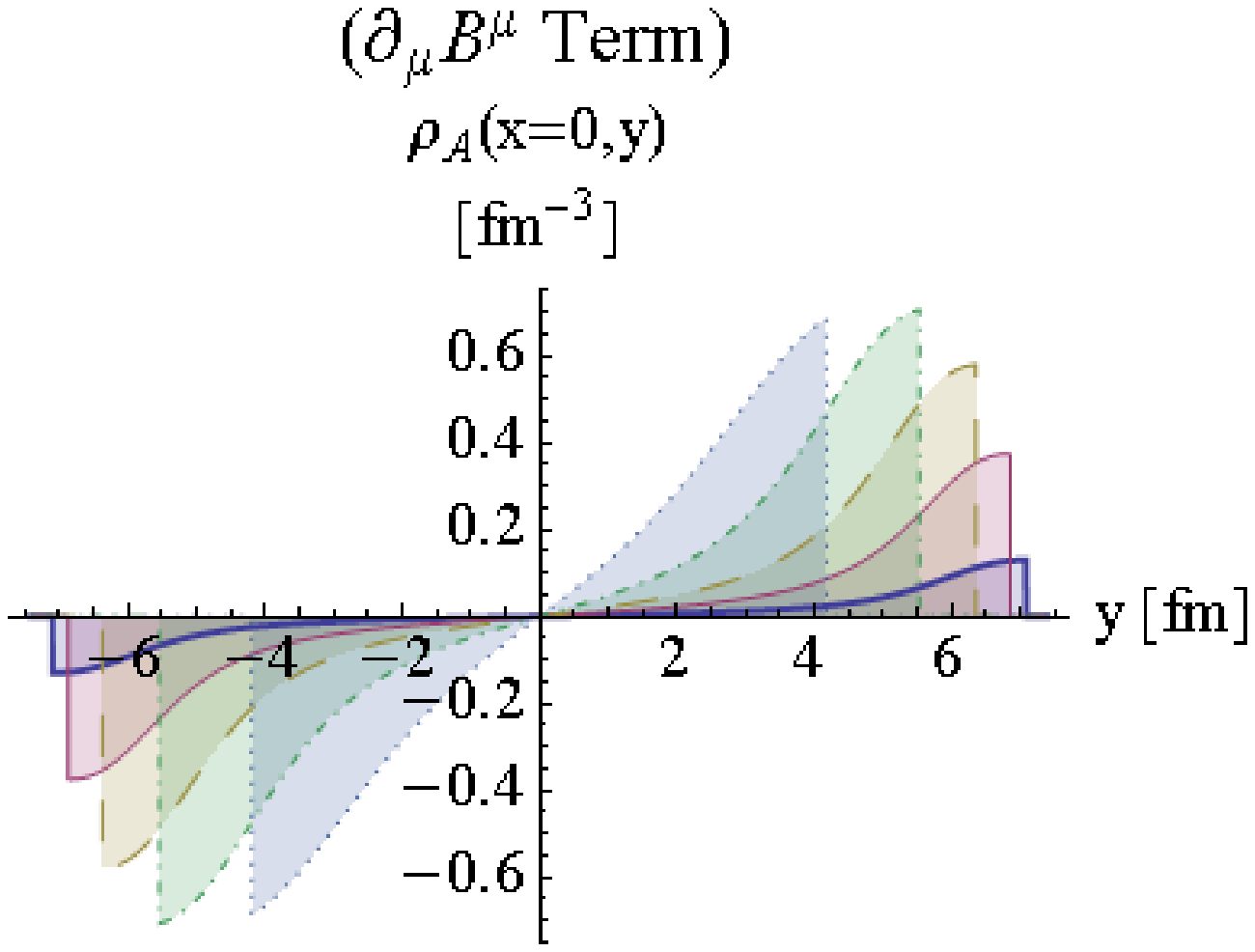}\includegraphics[width=5cm]{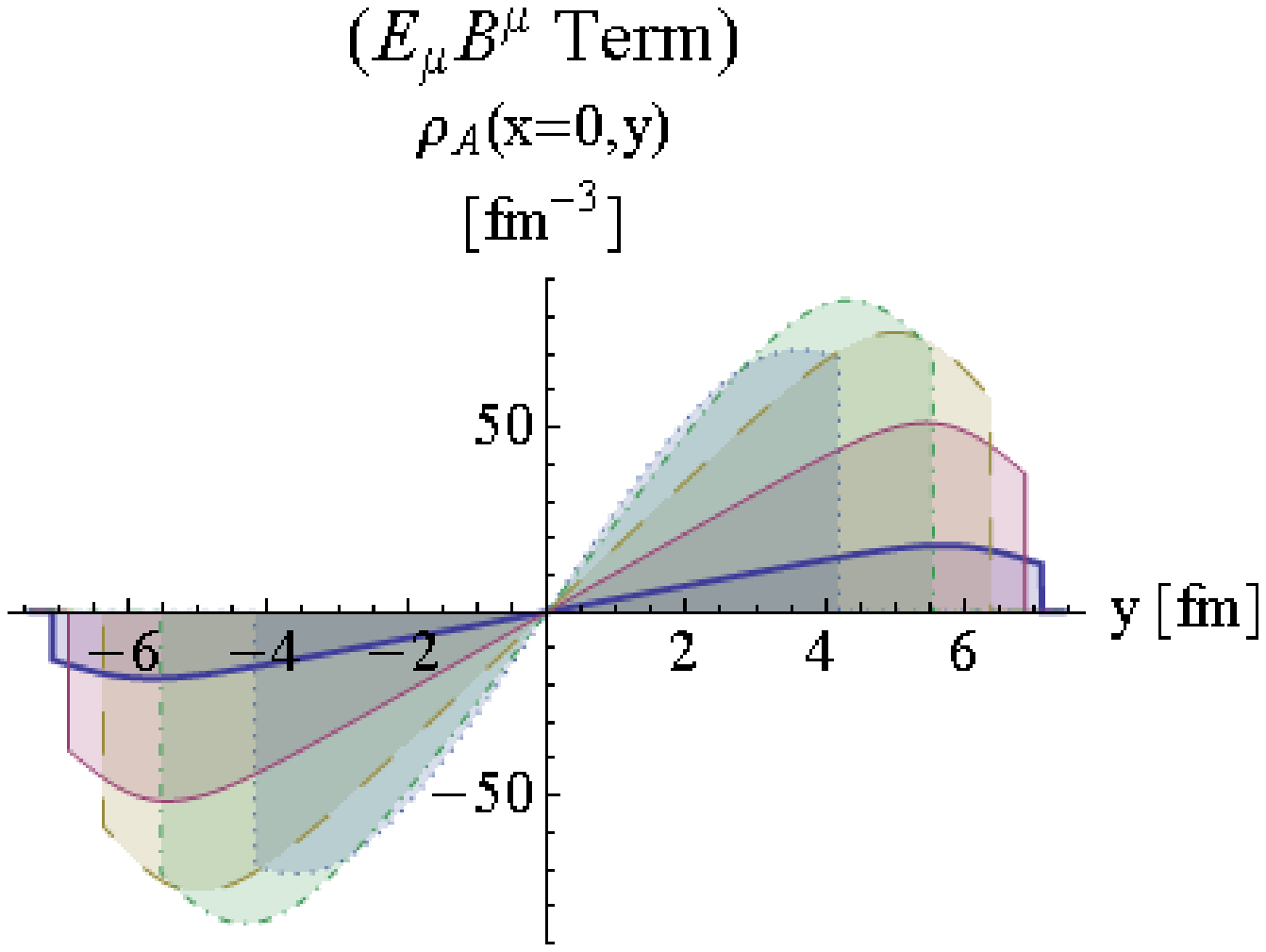}
\caption{The axial density for $t=t_0+\Delta t$, $x=0$ for various impact parameters.\label{fig_rhoAx0}}}

\FIGURE
{\includegraphics[width=5cm]{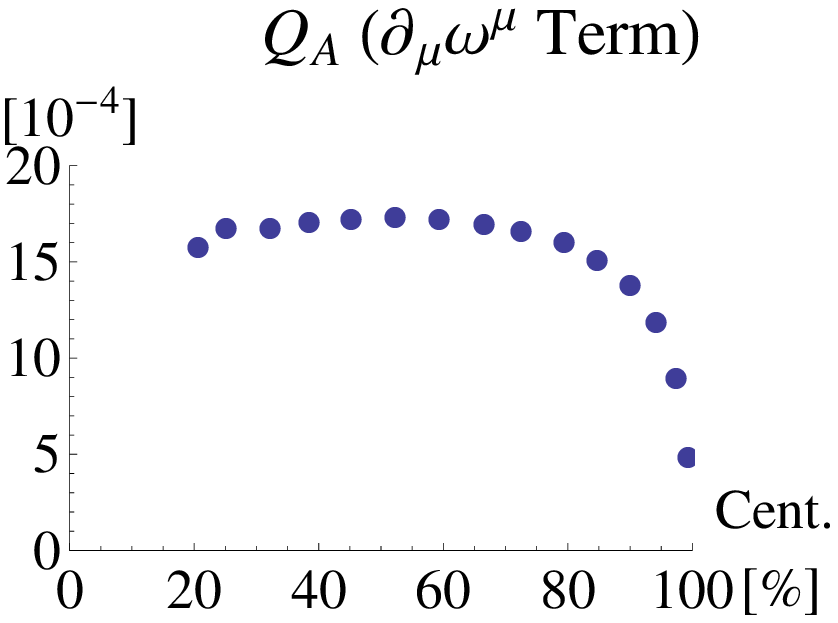}\includegraphics[width=5cm]{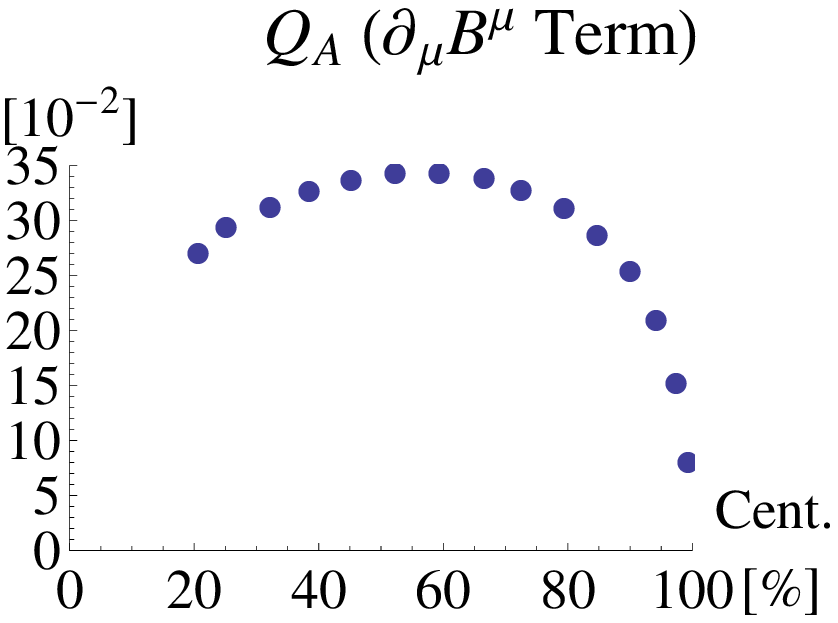}\includegraphics[width=5cm]{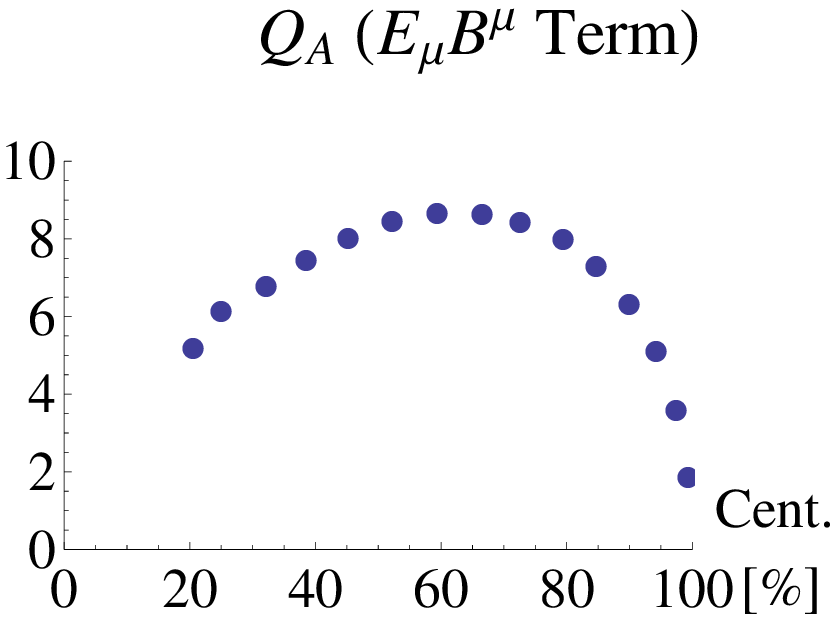}
\caption{The axial charge in the upper cone at $t=t_0+\Delta t$ as a function of centrality. The cone is centered along the rotation axis, with angular radius of $\Delta \theta_{cone}=60^0$.
The centrality is measured by the number of participants estimated using the Glauber model.
\label{fig_intrhoA}}}

\FIGURE
{\includegraphics[width=5cm]{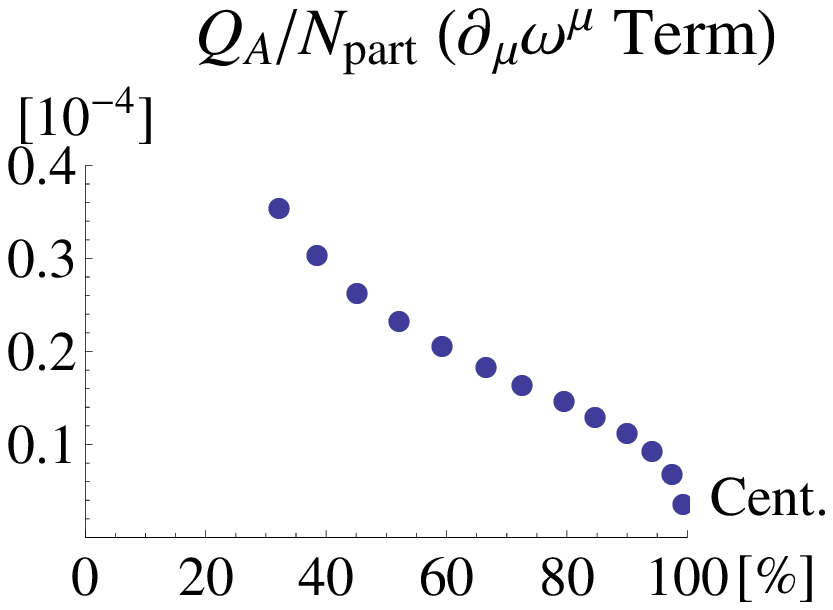}\includegraphics[width=5cm]{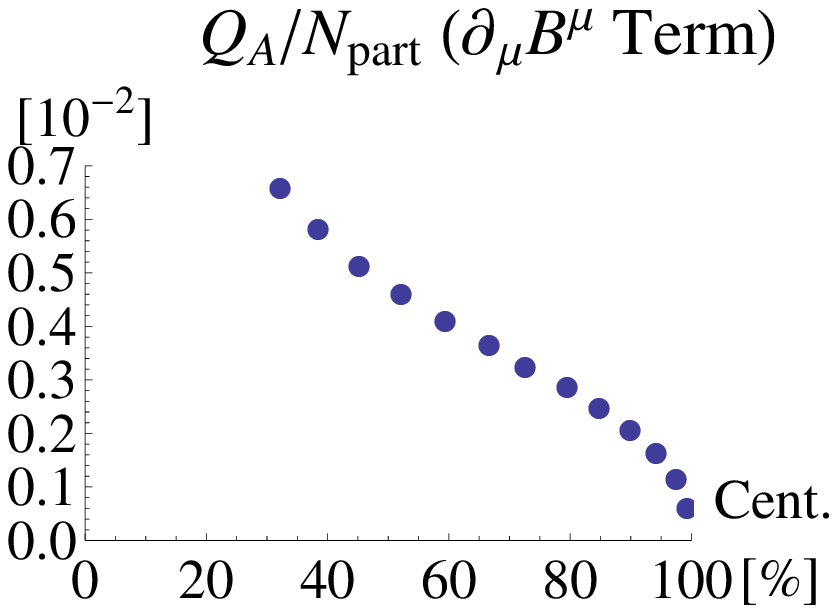}\includegraphics[width=5cm]{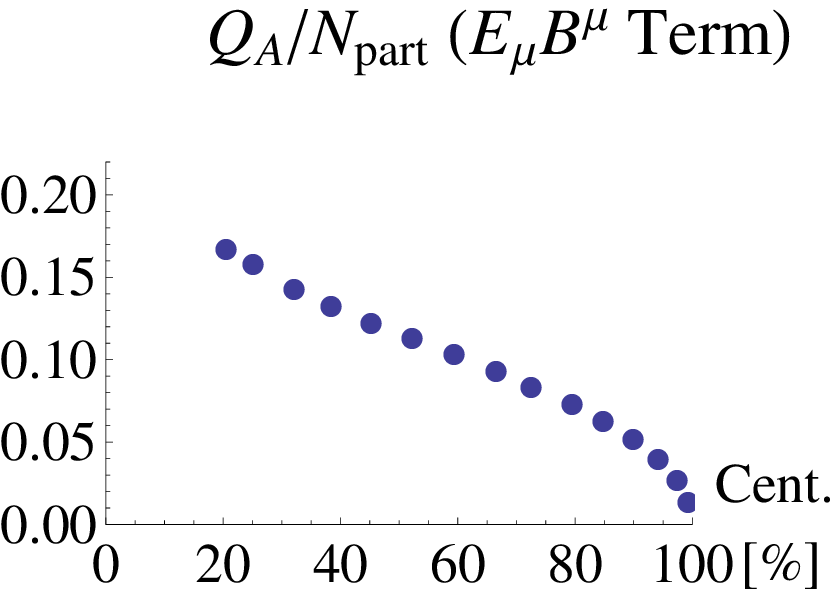}
\caption{The axial charge in the cone at $t=t_0+\Delta t$ as a function of centrality, divided by the number of participants. The cone is centered along the rotation axis, with angular radius of $\Delta \theta_{cone}=60^0$.
\label{fig_intrhoA_np}}}

\FIGURE{\includegraphics[height=4cm]{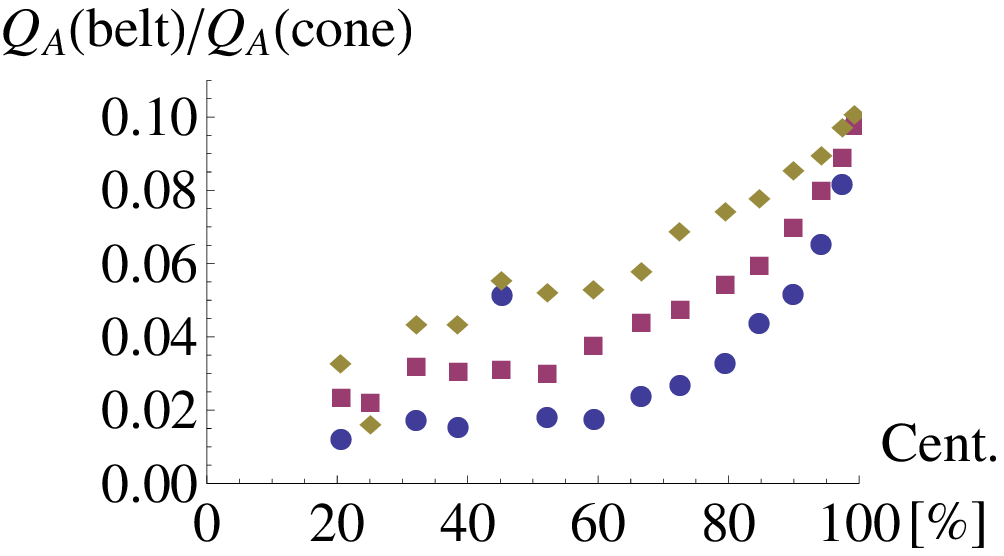}\includegraphics[height=4cm]{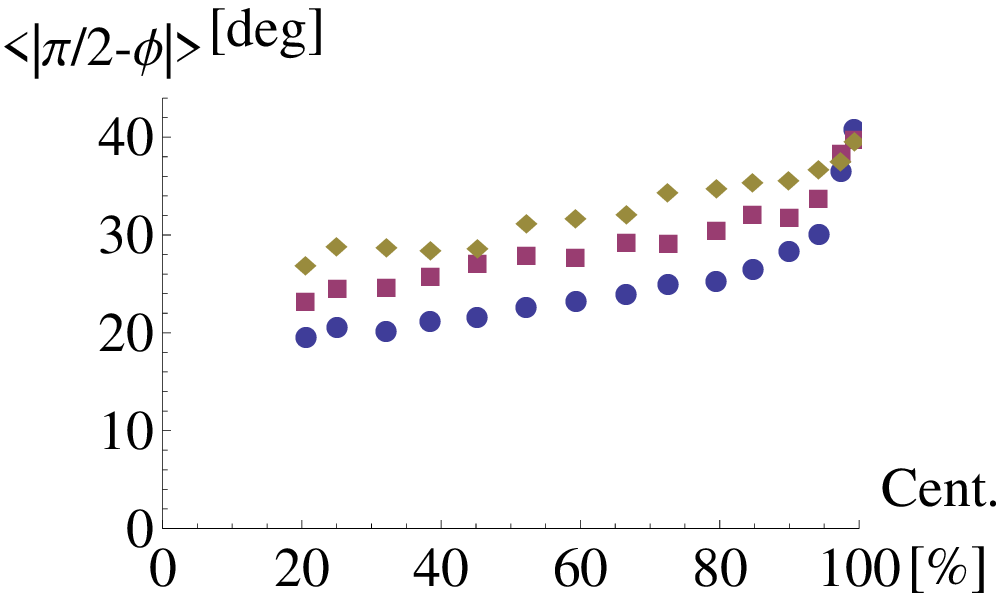}\includegraphics[height=4cm]{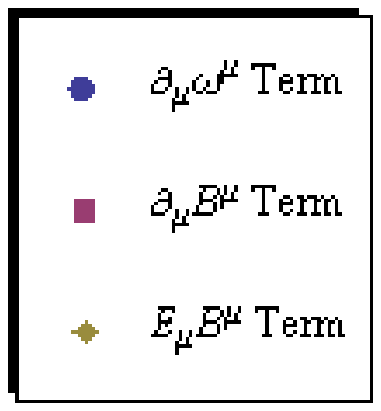}
\caption{The left plot show the ratio between the axial charge in the belt and in the cone with $\Delta \theta_{cone}=60^0$ and $\Delta \theta_{belt}=30^0$ (see fig. \ref{fig_prediction}). The plot on the right shows the second moment of the angular distribution defined in \ref{eq_pi_2_phi}.\label{fig_pi_2_phi}}}

\subsection{Signal Detection}
\label{sec_detection}
The signal discussed in section \ref{sec_signal} is proportional to the axial charge in a cone covered by the
angle $\Omega$. According to the results of section \ref{sec_numerical}, the angles with largest axial charge are centered along the angular momentum axis. According to figure \ref{fig_pi_2_phi}, most of the axial charge is located within a cone of angular radius $\Delta\theta_{cone}\approx30-40^0$.
Plot \ref{fig_rhoA} also demonstrates that matter scattered in the $x$ direction passes through a region of
zero axial charge, and should have regular production ratios.
The scattering into the belt around the equator, with width $\Delta \theta_{belt}$ can therefore be used to
measure $N_\Delta(\Omega_{(Q_A=0)})$.
The two relevant regions (the one with maximal axial charge and the one with zero axial charge) were shown in figure \ref{fig_prediction}.

The numerical computation described here can only provide us with the qualitative features of the effect under the linear evolution approximation. It does not give us the precise axial charge distribution at the moment of freeze-out.
This means that it is not clear at this stage of the analysis what are the
values of $\Delta \theta_{cone}$ and $\Delta \theta_{belt}$ which will give the most significant signal.

Another difficulty in the search for this signature is the fact that unflavored high-spin hadrons
are very short lived, and decay before they reach the detector.
As a solution we suggest to focus on narrower high-spin hadrons such as $\Omega^-$ (see e.g. \cite{Timmins:2009vy}).
The angular distribution of the production of $\Omega^-$ can be affected by charge and strangeness effects.
In order to isolate the spin-dependent effects, it is possible to use $\Xi^-$ as the representative of the spin $\half$ baryons. This choice of hadrons will decrease the statistics, but will also reduce the theoretical uncertainties.
$J/\psi$ is yet another narrow high-spin hadron which could be used as a probe for this signal, but in the case of the charm quark, it is not obvious whether the zero mass limit is valid.

The quantity we therefore propose to measure and compare to the plot is
\bea
\left[\frac{N_{\Omega^-}(cone)/N_{\Omega^-}(belt)}
{N_{\Xi^-}(cone)/N_{\Xi^-}(belt)}-1\right]
\eea
as a function of centrality, for various collision energies. As discussed above, this quantity is expected to be proportional to $\frac{Q_A}{N_{part}}$, and we therefore expect to find the functional behavior seen in plot \ref{fig_intrhoA_np}, with no dependence on collision energy.

\section{Discussion}

We presented a proposal for detecting effects of triangle anomalies in heavy ion collisions.
It is curious that such subtle quantum effects might be revealed in a collective fluid-like motion. 
Several parts of the presented analysis require careful study. The dynamics of phase transitions and the nature of
the colored currents in the fluid require a solid theoretical understanding. 
Of particular importance is a precise estimate of the magnitude of the effect, which requires
a numerical solution to the equations beyond the linear approximation. 
Several important assumptions need clarification:
\bi
\item
We assumed that in the deconfined phase the gluon field and its contribution to the axial current
can be estimated assuming a Coulomb interaction. 
\item
The estimate of the chemical potentials and the assumption that they are coordinate independent require a better study.
\item
The distribution of the strangeness current can also be centrality dependent. We assumed that comparing  $\Omega^-$ ($sss$) production with $\Xi^-$ ($ssd$) production can take this effect into account, and isolate the spin dependence, but this needs further study.

\item

In the current theoretical framework the freeze-out is successfully modeled by a sharp transition from a hydrodynamics description of deconfined QCD fluid with chiral symmetry restoration into a kinetic theory of hadron gas. However, if chiral symmetry breaking occurs before freeze-out then the enhanced spin alignment discussed here might be washed out before the quarks bind into hadrons. The study of this possibility requires a better understanding of the phase transition process.
 
\item
The assumption that the fermion number density is proportional to the number of participating nucleons requires
a stronger theoretical justification. The proportionality factor is a crucial ingredient in the analysis.
\item
In the analysis of section \ref{sec_signal} the volume over which the integration is performed was not specified in details. Since the process of freeze-out is not understood,
it is not clear what is the relevant volume. We performed the integral over the entire region of non-zero axial charge. Taking the integral over smaller regions should increase the effect, because the axial charge
is concentrated in regions of low participant density.
\item
We neglected the contribution of hadronization of gluons in the QCD fluid.
\ei
The last three items can be studied given a full numerical solution to the hydrodynamics equations,
and a modification of the freeze-out process computation.
In the full analysis, the number of particles produced during freeze-out is determined by the
off-equilibrium distribution functions\cite{Luzum:2008cw}
\bea
f_i(x^\mu,p^\mu)=g_i\exp(p_\mu u^\mu/T)\left[1+\frac{p_\mu p_\nu \tau^{\mu\nu}}{2T^2(\epsilon + p)}\right]
\eea
in which $g_i$ is the degeneracy of the $i$'th particle species, and $\tau^{\mu\nu}$ is the dissipative term discussed in (\ref{eq_rel_hydro}).
In order to take into account the effect of the axial charge we suggest the following modification:
\bea
f_i(x^\mu,p^\mu)&=&g_i\exp(p_\mu u^\mu/T)\left[1+\frac{p_\mu p_\nu \tau^{\mu\nu}}{2T^2(\epsilon + p)}
-a\frac{|\rho_A|u^\mu p_\mu}{(\epsilon + p)}\right]\nl
f_i^*(x^\mu,p^\mu)&=&g_i^*\exp(p_\mu u^\mu/T)\left[1+\frac{p_\mu p_\nu \tau^{\mu\nu}}{2T^2(\epsilon + p)}
+a\frac{\sum_j g_j}{\sum_kg_k^*}\frac{|\rho_A|u^\mu p_\mu}{(\epsilon + p)}\right]~,
\eea
where a quantity marked by $^*$ refers to spin excited hadrons (otherwise it refers to a ground state hadron),
the sums run over all produced particle species, and $a$ is a parameter that can be extracted from the data.
This phenomenological modification generates the effect discussed above, while keeping the total number of produced particles fixed. The results can be translated into a cosine expansion of the scattering cross section
\bea
\frac{dN_i}{d\phi}\propto(1+2v_{2i}\cos(2\phi))
\eea
and the expected result is a suppression of the $v_{2i}$ parameter for spin-excited hadrons.

\acknowledgments
We would like to thank A. Casher, M. Karliner, E. Kiritsis and M. Lublinsky for valuable discussions.
The work is supported in part by the Israeli
Science Foundation center of excellence, by the Deutsch-Israelische
Projektkooperation (DIP), by the US-Israel Binational Science
Foundation (BSF), and by the German-Israeli Foundation (GIF).

\end{document}